\documentclass[letterpaper,useAMS,usenatbib,usegraphicx]{mn2e}
\usepackage{times}

\newif\ifAMStwofonts
\AMStwofontstrue

\newcommand{\simlt}{\lower.5ex\hbox{$\; \buildrel < \over \sim \;$}}
\newcommand{\simgt}{\lower.5ex\hbox{$\; \buildrel > \over \sim \;$}}
\newcommand{\be}{\begin{equation}}
\newcommand{\ba}{\begin{eqnarray}}
\newcommand{\ee}{\end{equation}}
\newcommand{\ea}{\end{eqnarray}}

\title[Decoding the spectra of SDSS early-type galaxies]
{Decoding the spectra of SDSS early-type galaxies:\\
New indicators of age and recent star formation}
\author[B. Rogers et al.]
{Ben Rogers$^1$, Ignacio Ferreras$^{1,2}$\thanks{E-mail: ferreras@star.ucl.ac.uk},
Ofer Lahav$^2$, Mariangela Bernardi$^3$,
\newauthor
Sugata Kaviraj$^4$ and Sukyoung K. Yi$^5$\\
$^1$ Department of Physics, King's College London, Strand, London WC2R 6LS\\
$^2$ Department of Physics and Astronomy, University College London,
  Gower St. London WC1E 6BT\\
$^3$ Department of Physics and Astronomy, University of Pennsylvania, USA\\
$^4$ Astronomy group, The Denys Wilkinson Building, Keble Road, Oxford, OX1 3RH\\
$^5$ Centre for Space Astrophysics, Yonsei University, Seoul, South Korea\\
}

\begin{document}
\date{Accepted for publication in MNRAS, August 30, 2007.}
\pagerange{\pageref{firstpage}--\pageref{lastpage}} \pubyear{2007}
\maketitle
\label{firstpage}

\begin{abstract}
We apply Principal Component Analysis (PCA) to a sample of early-type
galaxies from the Sloan Digital Sky Survey (SDSS) in order to infer
differences in their star formation histories from their unresolved
stellar populations. We select a $z<0.1$ volume-limited sample
comprising $\sim 7000$ early-type galaxies from SDSS/Data Release~4.
Out of the first few principal components (PC), we study four which
give information about stellar populations and velocity dispersion. We
construct two parameters ($\eta$ and $\zeta$) as linear combinations
of PC1 and PC2. The four components can be presented as ``optimal
filters'' to explore in detail the properties of the underlying
stellar populations.  By comparing various photo-spectroscopic
observables -- including NUV photometry from GALEX -- we find $\zeta$
to be most sensitive to recent episodes of star formation, and $\eta$
to be strongly dependent on the average age of the stellar
populations. Both $\eta$ and $\zeta$ also depend on metallicity. We
apply these optimal filters to composite spectra assembled by Bernardi
et al. The distribution of the $\eta$ component of the composites
appear to be indistinguishable between high and low density regions,
whereas the distribution of $\zeta$ parameters have a significant skew
towards  lower values for galaxies in low density regions.
This result suggests that galaxies in lower density environments
are {\sl less} likely to present weak episodes of recent star formation. In
contrast, a significant number of galaxies from our high density
subsample -- which includes clusters (both outer regions and centres)
and groups -- underwent small but detectable recent star formation at high
metallicity, in agreement with recent estimates targeting elliptical galaxies
in Hickson Compact Groups and in the field (Ferreras et al.).
\end{abstract}


\section{Introduction}

Observational studies of galaxy formation follow a two-pronged
approach.  On the one hand dynamical studies relate to 
the past mass assembly history of galaxies. We believe this
history is strongly driven by the ``bottom-up'' hierarchical merging
scenario as predicted by the $\Lambda$CDM paradigm by which massive
halos form from the progressive mergers of smaller systems. Massive
cosmological simulations computed within the $\Lambda$CDM framework
give large scale properties in agreement with available surveys
(Springel et al. 2005).  

On the other hand, the star formation history
tracks the transformation of gas into stars, a process controlled both
by the merging structure described above as well as by the highly
non-linear physics of star formation. Even full-fledged cosmological
models of structure formation have to rely on simple prescriptions of
star formation as the dynamical range required for a consistent
treatment is far beyond the current capabilities of numerical
simulations (see e.g. De Lucia et al. 2006). The complexity of star
formation and its possible feedback mechanisms represent our current
bottleneck in numerical studies of galaxy formation and justify
alternative approaches aimed at extracting information about the star
formation history from photo-spectroscopic properties of the
unresolved stellar populations.

The pioneering work of Tinsley (see e.g. Tinsley 1980) was based on a
comparison of well-defined models of galaxy formation including
stellar evolution with photometric data in order to assess the ages
and metallicities of galaxies. This work was continued by more
detailed population synthesis models (e.g. Bruzual \& Charlot
2003). The age-metallicity degeneracy (Worthey 1994) by which age and
metallicity effects generate similar photo-spectroscopic trends
represent the main hurdle constraining the star formation history
of galaxies. This problem is more severe when dealing with the old
stellar populations found in early-type galaxies.  Subsequent work
targeting age-sensitive absorption lines (Worthey \& Ottaviani 1997;
Vazdekis \& Arimoto 1999; Trager et al. 2000) is arguably the best approach to
quantifying differences between stellar populations, although caveats
exist (see e.g. Prochaska et al. 2007).

Other approaches aim at using as much information as possible (namely
the whole spectral energy distribution), but comparisons are hindered
by the sheer volume of parameter space to be explored. Ingenious
approaches (Heavens, Jimenez \& Lahav 2000; Panter, Heavens \& Jimenez
2003; Ocvirk et al. 2006) have been proven useful to decipher the star
formation histories of galaxies but there are uncertainties as to the
accuracy of the parameters extracted by the models. The effect of
systematics -- such as flux calibration errors -- can translate into
large uncertainties on the estimates of the stellar ages.

Multivariate techniques such as Principal Component Analysis or
Independent Component Analysis (Hyv\"arinen, Karhunen \& Oja 2001)
follow a different approach. Instead of a direct comparison between
data and models, the observations are considered to be the sole source
of information. These data are classified or rearranged in a way that
maximises the amount of information (in the sense of variance) carried
by the spectra. The models are used ``a posteriori'' to put the
physics back in. The advantage of this approach is that trends seen on
the rearranged spectra (i.e. the principal components, independent
components, hidden variables, etc) are robust and model-independent.

PCA has been previously applied to various sets of astrophysical
spectra; from stars (Deeming 1964) to active galactic nuclei (Francis,
Hewett, Foltz \& Chaffee 1992). Most notably, PCA has allowed the
classification of galaxy spectra from surveys such as 2dF (Madgwick et
al. 2003) and SDSS (Yip et al. 2004). Our approach is different. We
use PCA not as a classification tool for a wide range of spectra, but
as a method to extract minute differences in the stellar populations
of an otherwise highly homogeneous sample. Our work is a continuation of 
the pioneering studies of Faber (1973) and more recent work on SDSS
galaxies by Eisenstein et al. (2003).

These multivariate techniques have been mainly applied to the
classification or compression of spectra (e.g. Yip et
al. 2004). Recently, these techniques have been pursued on galaxy data
with another purpose: to extract differences in the underlying stellar
populations in order to determine the star formation histories. For
that purpose, elliptical galaxies represent one of the best
systems. These galaxies pose a well-known problem: dominated by old
stellar populations -- for which the time evolution is very slow --
they present a strong age-metallicity degeneracy (Ferreras, Charlot \&
Silk 1999).

At the same time, these galaxies represent one of the best places to
constrain our knowledge of galaxy assembly. Massive ellipticals are
dominated by old stars, suggesting an old, brief and intense period of
star formation (the so-called monolithic collapse). However, our
understanding of structure formation requires these galaxies to be
assembled later on via massive mergers.  Even though semi-analytic
models improve and give reasonable answers to this problem 
(Kaviraj et al. 2005; De Lucia et al. 2006; Kaviraj et al. 2007), 
the models do not have enough resolution to target
important issues related to star formation.  For instance, the
alpha-enhanced values measured in massive ellipticals are still a
matter of controversy in the light of hierarchical models (Thomas
1999).  The star formation prescriptions used by semi-analytic models
(see e.g. Croton et al. 2006) are clearly insufficient to account for
the effects that cause these overabundances.  Furthermore, the effect
of environment on the stellar populations is also an important issue:
hierarchical models predict significant differences in the star
formation histories of galaxies with respect to local density. In
essence, galaxies in denser regions form earlier (Sheth \& Tormen
2004). However, detailed spectroscopic observations do not show large
differences in the stellar populations of field and cluster
ellipticals (Bernardi et al. 1998; 2006).

Recent studies applying various multivariate techniques (Kab\'an, Nolan
\& Raychaudhury 2005; Ferreras et al. 2006; Nolan, Raychaudhury \&
Kab\'an 2007) to the spectra of early-type galaxies have started to give
very valuable insight into the underlying stellar populations of these
systems. This information is gathered in a total independent way to
model-oriented techniques. Hence, this methodology of tackling the star formation
history of unresolved stellar populations is becoming a valuable
complement to studies based on equivalent widths.

\section{The sample}

Our sample is extracted from the catalogue of Bernardi et al. (2006)
which comprises $\sim $42,000 elliptical and lenticular galaxies
selected from the Sloan Digital Sky Survey (York et al. 2000). From
this starting catalogue, we select a volume-limited sample to redshift
$z<0.1$. We examine a plot of redshift versus absolute magnitude
(in the $r$ band) of the original sample to determine the cut to be imposed on the
absolute magnitude for a $z<0.1$ sample. We find this cut to be at 
$M_r<-20.7$. Furthermore, we impose a limit of S/N$\geq 15$ to avoid
noisy spectra. The cut in S/N results in a small fraction of galaxies
at the faint end to be rejected but within Poisson error bars we 
tested that no significant bias is introduced.

The preliminary sample comprises $9192$ galaxies. These
galaxies are compared with the luminosity function of early-type
galaxies measured in SDSS (Nakamura et al. 2003) to
find that a further cut at $M_r\leq -21$ should be imposed to obtain a
volume-limited sample. Our final sample thereby comprises $7148$
galaxies out to $z<0.1$ and brighter than $M_r\leq -21$. The
characteristic luminosity from the Schechter function fit to SDSS
early-type galaxies is $M_r(\star )=-21.6$. Hence, our galaxies
correspond to $L_r\geq 0.6L_{r,\star}$.  The average redshift of the
sample is $\langle z\rangle =0.075$.

The SEDs were extracted from Data Release 5 (Adelman-McCarthy et al. 2007). 
The spectra were dereddened -- using the Fitzpatrick (1999) R$_v=3.1$ Galactic
extinction curve, taking the reddening values from the maps of
Schlegel, Finkbeiner \& Davis (1998) -- and deredshifted using a
linear interpolation algorithm and the redshift estimates supplied by
SDSS.  Finally, the SEDs were normalised to the same average flux
across the 6000-6500\AA\ wavelength range.

\section{Principal Component Analysis as a tool to rearrange galactic spectra}

Principal Component Analysis (PCA) is a multivariate technique aimed
at reducing the dimensionality of a data set in order to decrease the
complexity of the analysis. Each member of the sample (a galaxy SED in our
case) is defined by an ``information vector'', an N-tuple of numbers 
given by the flux measured at a set of wavelengths: $\{\Phi (\lambda_i)\}, 
i=\{1,2,\cdots N\}$. PCA consists of decorrelating sets of vectors
by performing rotations in the n-dimensional parameter space spanned by
the wavelength bins.  The final result is a diagonal covariance matrix, with
the eigenvalues representing the amount of information (in the sense of
variance) stored in each eigenvector, which is called a principal component.
Rearranging the principal components in decreasing order allows us to
determine the main ones that contribute to determine the information
vectors of the input spectra.

Details of the PCA technique for our specific problem can be found
elsewhere (Ferreras et al. 2006).  Notice that in our methodology we
do not subtract the mean from each of the input vectors, i.e. we do
not eliminate the average spectrum. This implies the matrix we are
diagonalising is not the covariance matrix. As stated in Ferreras et
al. (2006), subtracting the mean removes information from PCA,
specifically about the shape of the continuum, which would be
preferable to keep within the analysis. Since the sample used here is
restricted to early-type galaxies, for which only small differences
are apparent between spectra, the mean and first eigenvector, $e_1$
should be similar. PCA done with such mean subtraction indeed confirms
this idea. A more orthodox ``mean-subtracted'' PCA is well justified
in cases such as the classification of 2dF spectra (Madgwick et
al. 2003), where the subtraction of the mean alleviates the strong
uncertainties in their flux calibration. Nevertheless we performed PCA
both with and without mean-subtraction on our sample to find no
significant difference in the outcome. The net effect is, roughly, a
shift in the rank of the principal components, so that $e_2$ becomes
$e_1$ in the mean-subtracted version, etc.

We apply PCA to different ``flavours'' of the observed data. In order
to assess the robustness of the extraction of the components we
compare the projections of the galaxy spectra on to the principal
components obtained in several different ways. We present the spectra
to PCA for several wavelength ranges: 3850--4150\AA (just encompassing
the 4000\AA\ break); 3850--4400\AA (including information from the
G-band); 3850--5000\AA\ (our fiducial wavelength range); and finally
continuum-subtracted spectra (using a 200\AA\ boxcar median filter to
determine the continuum) over the 3850--5000\AA\ range. The results
from these different sets of spectra allow us to assess whether flux
calibration errors can introduce a systematic effect.  We decided not
to extend our analysis to wavelengths redder than rest-frame 
$\lambda=5000$\AA\ as many of the SDSS spectra suffer from a poorly subtracted
night sky line (OI $\lambda = 5577$\AA\ ). PCA is very sensitive to
outliers, which contribute excessively to the variance. Furthermore,
the region around the night sky line cannot be simply masked out as
the position of this line gets deredshifted along with the galaxy
spectra to different wavelengths depending on the redshift of each
galaxy. We tried various methods, including a replacement of the
region around the emission line with a normalized average of all the
spectra, but the results were not satisfactory.  In order to make our
results as robust as possible we decided not to tresspass the 5000\AA\
limit, which keeps the 5577\AA\ OI night sky line safely away from the
spectra of all of our $z<0.1$ galaxies.

Flux calibration errors can have an effect on the projected components.
Blind techniques such as PCA rely heavily on data which should have 
systematic effects kept under control. Data gathered from various 
sources can yield results directly related to differences in the
systematics of these sources. By choosing a homogeneous sample (SDSS DR5)
we minimise such effects. Furthermore, the SDSS
errors of flux calibration are stated to be limited to $\sim$3\%
(see www.sdss.org/dr4/algorithms/fluxcal). We find flux calibrations
of this order not to give any strong differences in the projected components
presented here.

\begin{table}
\caption{Weight of the first Principal Components}
\label{tab:flavs}
\begin{center}
\begin{tabular}{ccccc}
\hline\hline
Label & Spectral Range & $\lambda_1$(\%) & $\lambda_2$(\%) & $\lambda_3$(\%) \\
\hline
 1 & 3850--4150\AA\   & 99.183 & 0.072 & 0.019 \\
 2 & 3850--4400\AA\   & 99.425 & 0.055 & 0.015 \\
 3 & 3850--5000\AA\   & 99.693 & 0.067 & 0.006 \\
 4 & CS$^a$3850--5000\AA\ & 69.305 & 0.880 & 0.336 \\
\hline\hline
\end{tabular}
\end{center}
$^a$ CS = Continuum Subtracted.
\end{table}


Table~\ref{tab:flavs} shows the ``weight'' of the first few principal
components.  PCA applied to the full spectra gives a dominant first
principal component (a result of the high homogeneity of the stellar
populations in early-type galaxies).  The strength of this first
component increases as the wavelength range is increased.  In
contrast, the continuum subtracted spectra give a more spread out
distribution of weights among the first principal components. However,
the first component still dominates the variance.

\begin{figure*}
\includegraphics[width=15cm]{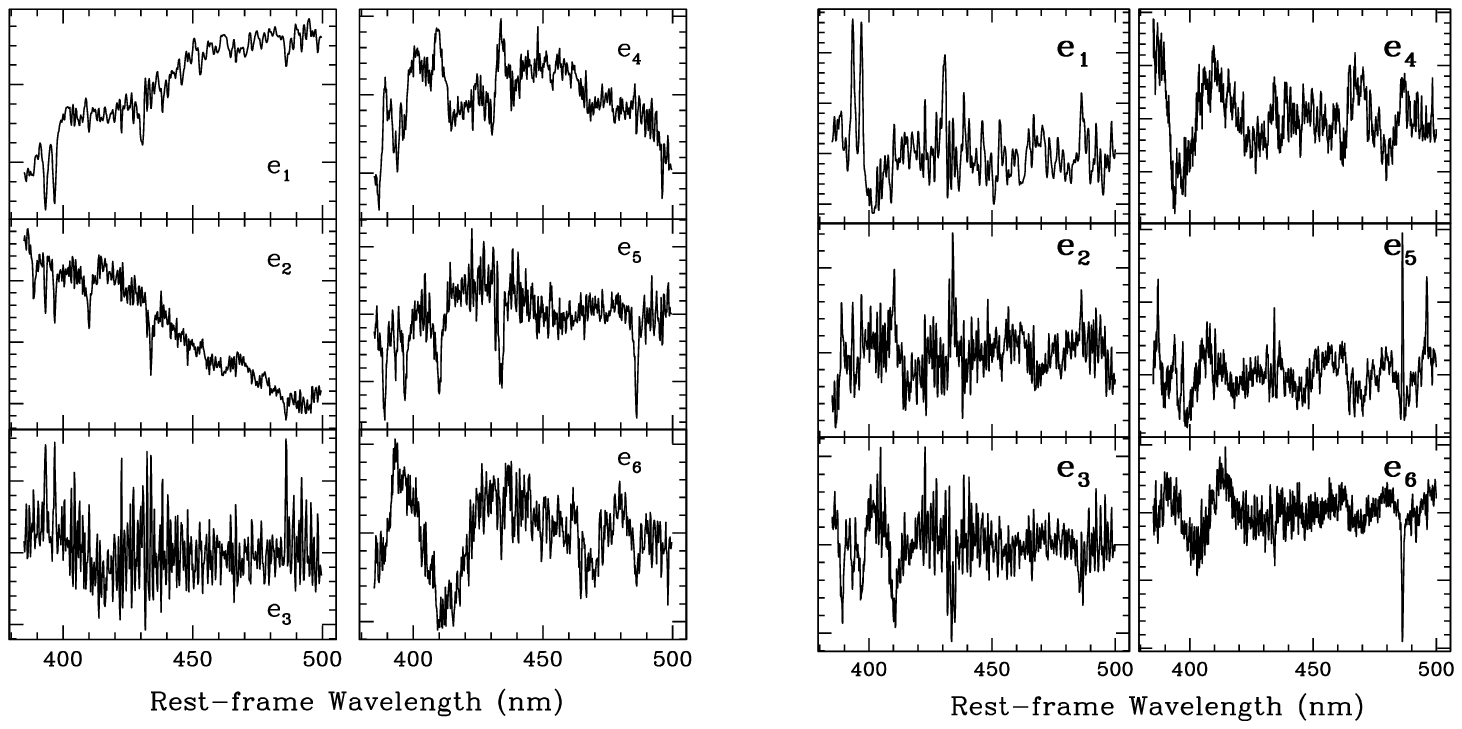}
\caption{The first six principal components using the full SED
({\sl left}) and the continuum subtracted SED ({\sl right}), over
the wavelength range 3850--5000\AA\ . These components correspond
to ``flavours'' 3 (left) and 4 (right) in table~\ref{tab:flavs}.
The projection (``dot product'')
of these eigenvectors ($e_1, e_2,\cdots$) with each galaxy is denoted 
throughout the paper as PC1, PC2, etc. 
}
\label{fig:PC_seds}
\end{figure*}

The first Principal Components obtained both with the full SEDs and
with the continuum subtracted ones are shown in
figure~\ref{fig:PC_seds}.  The first component shows the
characteristic features of old stellar populations, a prominent
4000\AA\ break and no Balmer absorption.  The calcium H and K lines
are also visible along with the G band, clear indicators of cool
stars. The second component can be physically attributable to a young
stellar population, with prominent Balmer absorption and a blue
continuum. Its low variance ($\sim 0.07$\%) is consistent with the
lack of young stars in early-type galaxies. 
Our first and second principal components are analogous to the
$a1$ and $a2$ components of Nolan et al. (2007), who used factor
analysis to decompose the spectral information from a sample of
early-type galaxies from SDSS/DR4.  Our third principal component is
notably noisier than the first two, but notice the next components do
not appear as noisy. We will show below that the third component
correlates strongly with the velocity dispersion, a property that
appears throughout the SED, hence its noisy appearance.  Of the higher
order components it is worth mentioning the fifth component, which
clearly presents the Balmer series. This component is remarkably
similar to the one obtained by applying PCA to a smaller sample of
early-type systems in the field and in Hickson Compact Groups
(Ferreras et al. 2006; Ferreras et al. 2006b).  However, the sample
studied here is several orders of magnitude larger and the Balmer
lines appear at a higher S/N ratio. The very low variance associated
to this component (around $0.002$\%) implies Balmer-strong spectra --
i.e. the so-called E+A or k+A galaxies -- are not common in a large,
low-redshift, volume-limited sample of early-type systems such as the
one presented here. The sample selection done by Bernardi et
al. (2006) do not bias in any way against the selection of this type
of galaxies, hence the low variance associated to this component
should reflect the weight of these post-starburst galaxies in the
census of $z<0.1$ early-type systems, in contrast with the population
of such systems at moderate and high-redshift (Tran et al. 2003).  We
should emphasize that the long-range structure seen in high order
components such as $e_4$ or $e_6$ is not physical, but an artifact
generated by the enforced orthogonality of the principal components.

The panels on the right hand side of figure~\ref{fig:PC_seds} show the principal
components for the continuum-subtracted spectra. The features are harder
to interpret, but the calcium H+K lines, the G band
and some of the Balmer absorption lines are evident.

\begin{figure*}
\includegraphics[width=15cm]{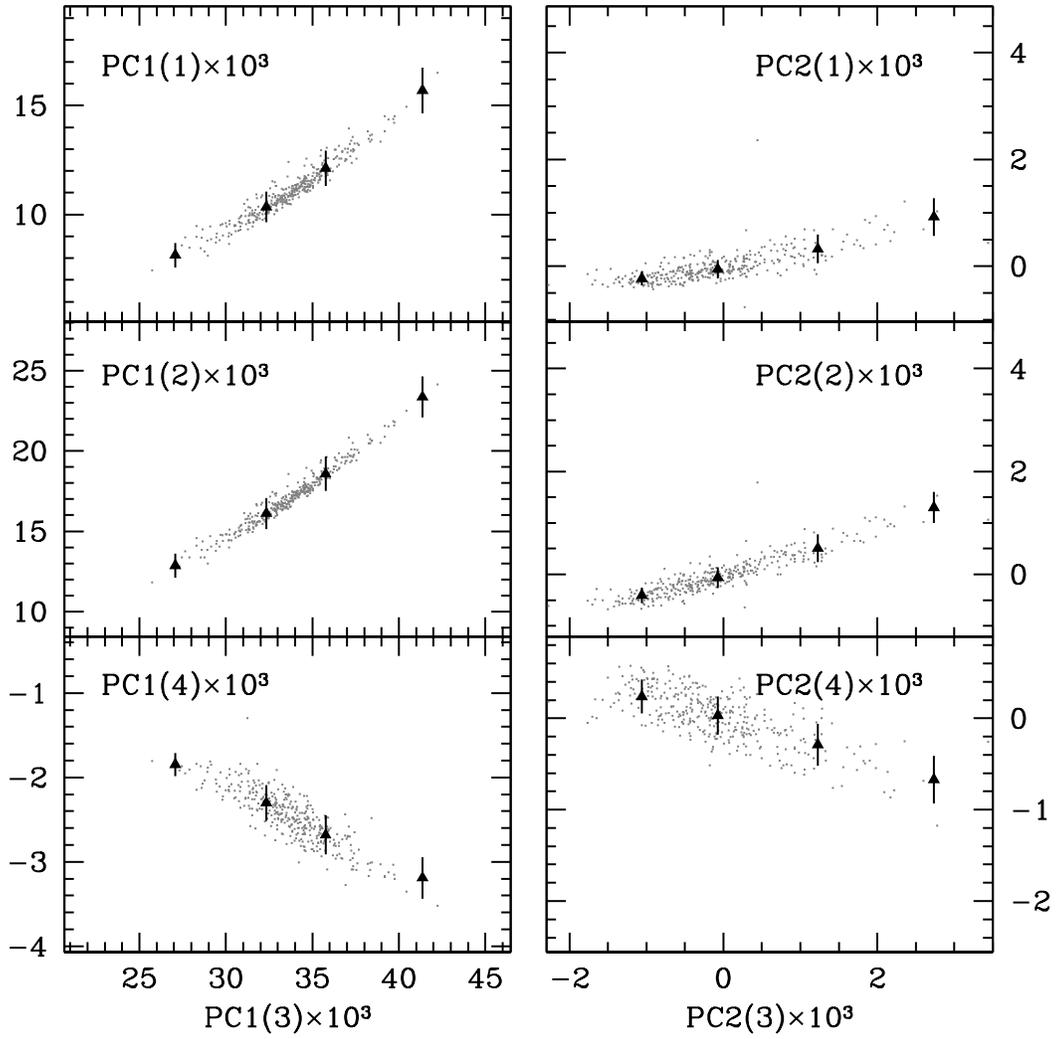}
\caption{Comparison between the projection of the sample
galaxies on the first two principal components for four different
``flavours'' of the spectra as explained in the text. The
indices in brackets correspond to those in table~\ref{tab:flavs}. The
black triangles and error bars are the average and median of
the samples binned in PC1(3) and PC2(3).}
\label{fig:flavs}
\end{figure*}

Figure~\ref{fig:flavs} gives the projections of the galaxies on the
first two principal components for different options of the spectral
range or for the continuum subtracted spectra (as labeled in
table~\ref{tab:flavs}).  The figure shows that these different
``flavours'' give very similar projections of the galaxy spectra on to
the principal components, i.e. the decomposition is fairly robust, and
the results do not change significantly even when the information from
the continuum is subtracted. Henceforth, we will use the data with the
maximum amount of information, namely the full SED over the wavelength
range 3850--5000\AA\ .

\subsection{Projected Components}

The principal components can be considered a set of basis vectors that
optimally filter the information hidden in the spectra. We can rearrange
the information hidden in the $N$ components $\{\Phi (\lambda_i)\}$, 
$i=\{1,2,\cdots N\}$ by
projecting the SEDs on to the principal components. This is equivalent
to a rotation in the $N$-dimensional parameter space spanned by the
wavelength sampling. 
\be 
{\rm PC1}_i =
\vec{\Phi}_i\cdot\vec{e}_1=\sum_{j=1}^N
\Phi_i(\lambda_j)e_1(\lambda_j),
\ee  
and analogously for PC2, PC3, etc. According to the sharply decreasing
eigenvalues, one would expect most of the information (in the sense of
variance) to stay in the first few principal components.

\begin{figure}
\begin{center}
\includegraphics[width=16cm]{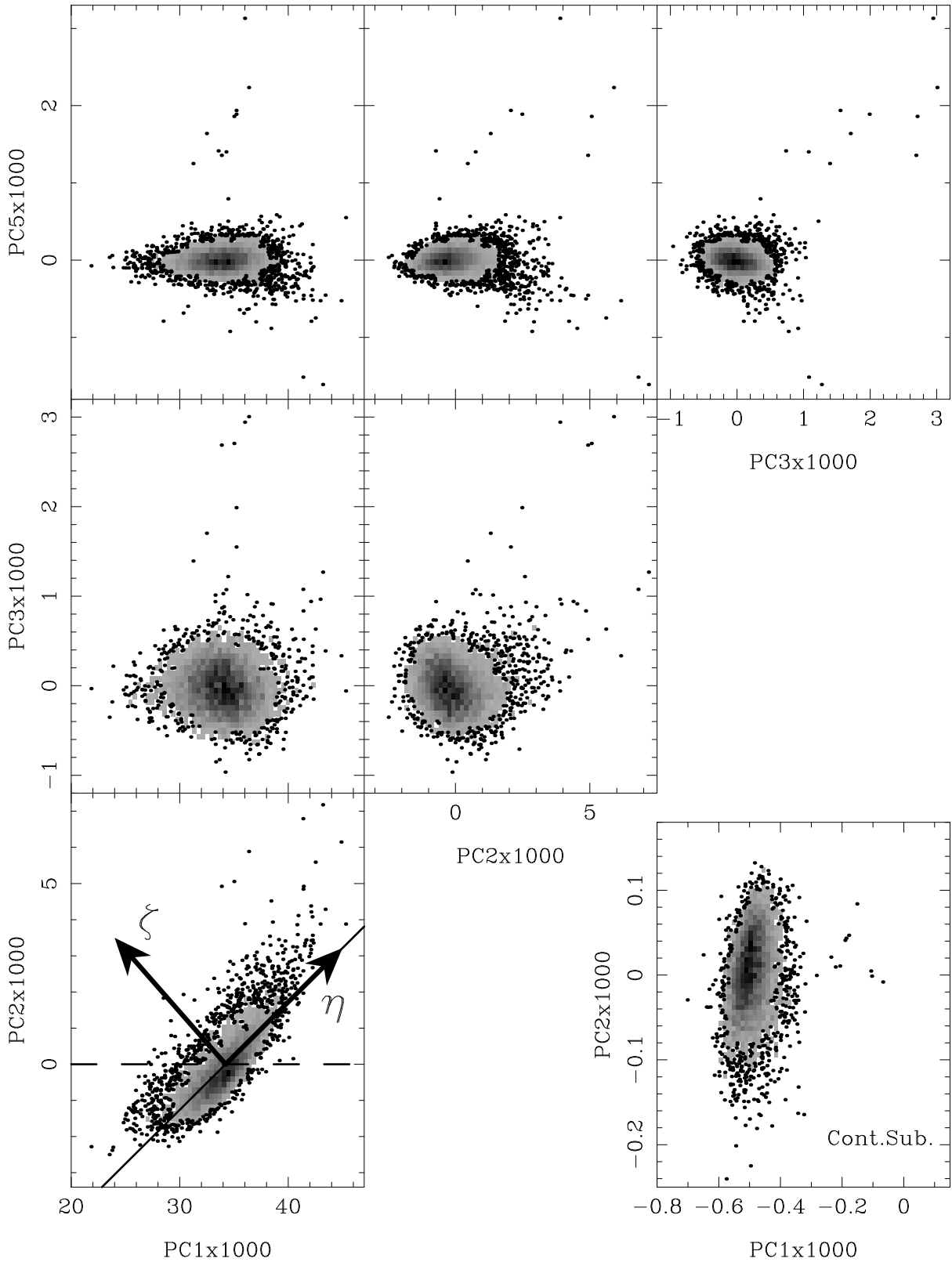}
\end{center}
\caption{Projection of the galaxy spectra on principal components
PC1, PC2, PC3 and PC5 (``flavour'' 3 in table~\ref{tab:flavs}). 
The greyscale maps the number density of
galaxies in PC-space. In the outer regions the greyscale is replaced
by dots representing individual galaxies. The bottom-right panel
gives the result for continuum subtracted spectra 
(``flavour'' 4 in table~\ref{tab:flavs}). $\eta$ and
$\zeta$ are linear combinations of PC1 and PC2 which further separate
the components.}
\label{fig:proj}
\end{figure}

Figure~\ref{fig:proj} shows the projection of the spectra of all
galaxies in our sample on components 1, 2, 3 and 5. Given the number
of galaxies to be plotted, we present the figure as a greyscale
corresponding to the density of galaxies. For those regions with lower
densities in PC parameter space, we replace the greyscale by dots
representing the projections of individual galaxies.  The projections
of the first two components, PC1 and PC2, show a correlation, present
both in the full SED (bottom left) as well as in the
continuum-subtracted one (bottom right). This result is consistent
with the analysis of the smaller sample comprising field and Hickson
Compact Group galaxies (Ferreras et al. 2006). We tested against a 
bias caused by S/N by comparing separate projections of the galaxies
with high and low S/N to find the same trend.
The projections of the
higher order components do not show any significant correlation with
PC1 or PC2, althogh we find the outliers in the PC1 vs PC5 diagram to be
galaxies with very prominent Balmer lines (i.e. post-starburst
galaxies). Notice the remarkably small fraction of such systems found
with high projections of PC5 in this large volume-limited sample.

The bottom-left panel of figure~\ref{fig:proj} also shows a simple
linear fit to the data. In the same spirit as the decomposition done
by Madgwick et al. (2003) on the PC1-PC2 plane spanned by their sample
of 2dF galaxies, we rotate the PC1-PC2 plane into two independent
components $\eta$ (the length of the projection along the straight
line that describes the fit) and $\zeta$ (the distance to the
line). As shown in Ferreras et al. (2006), such a definition allows us
to enhance the mapping between principal components and actual
physical parameters.

\begin{figure*}
\includegraphics[width=5in]{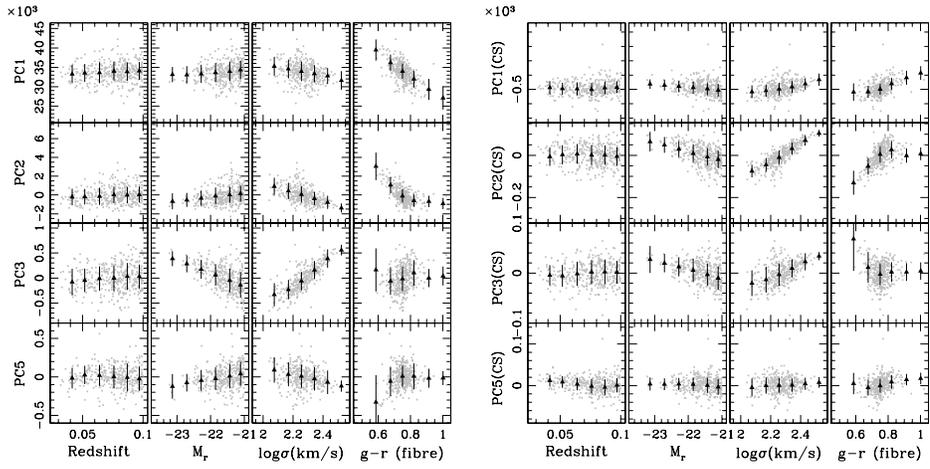}
\caption{Correlation of the galaxy projections with respect
to physical observables for the full spectra ({\sl left}) and the
continuum-subtracted SEDs ({\sl right}). The black triangles and error
bars give the average and standard deviation of the sample.}
\label{fig:physobs}
\end{figure*}

\begin{figure}
\begin{center}
\includegraphics[width=5in]{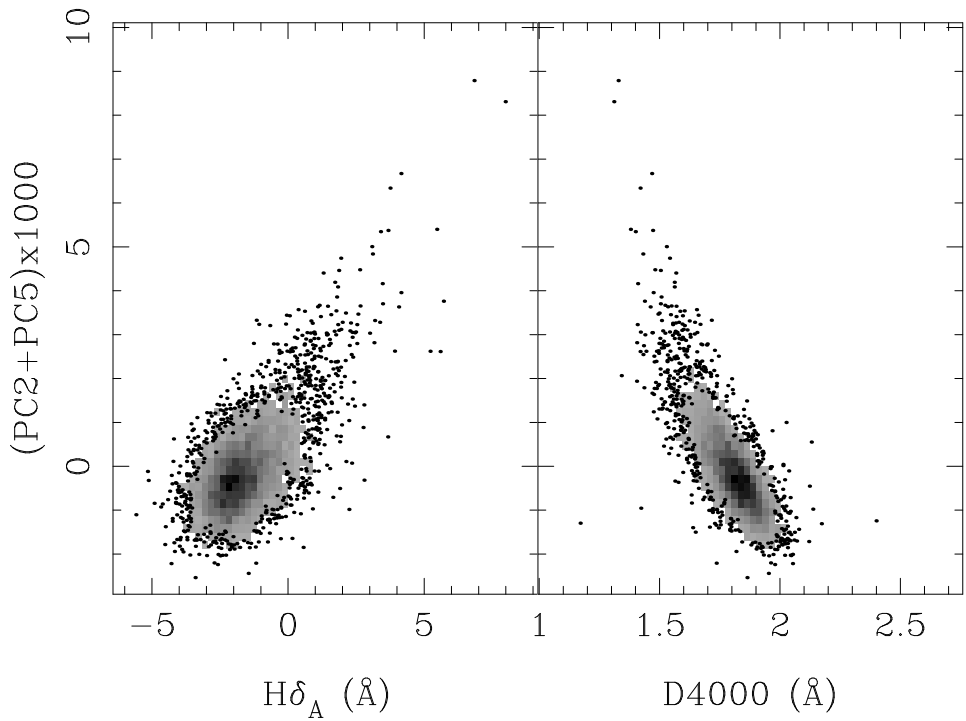}
\caption{
Correlation between a combination of PC2 and PC5
and the equivalent width of H$\delta_A$ 
({\sl left}, as defined in Worthey \& Ottaviani 1997);
and the strength of the 4000\AA\  break ({\sl right},
as defined in Balogh et al. 1999).}
\label{fig:EWs}
\end{center}
\end{figure}

Figure~\ref{fig:physobs} compares the projection of the principal
components with a set of physical observables such as redshift,
absolute magnitude, velocity dispersion or colour (as measured inside
the fibre used for the extraction of the spectra). There are two sets
of panels corresponding to the analysis of the full spectra (left) and
the continuum-subtracted version (right). Each figure shows in grey
the projections of the galaxy spectra on to the principal components,
as well as the binned average and variance. The figure illustrates the
clear correlation between PC1,PC2 and colour, to be expected in the
full spectra as the principal components correspond to a red and blue
spectrum, respectively. However, it is quite remarkable to find such a
strong correlation for the continuum-subtracted case, where this
colour information has been removed when presenting the spectra to
PCA. It is also worth mentioning the strong correlation found between
PC3 and velocity dispersion. A similar correlation appears in PC1 and
PC2, however we believe such is an indirect effect caused by the
well-known relation between velocity dispersion and colour (see
e.g. Bernardi 2003; 2005). No strong correlation is apparent with
PC5. Since PC5 is associated to Balmer absorption, one could again
interpret this result as the lack of important ``activity'' regarding
post-starburst spectra in the $z<0.1$ universe.
The correlation between projections PC2 and PC5 and a young stellar
component is also illustrated in figure~\ref{fig:EWs} where a
combination of both projections are compared with the equivalent
width of H$\delta_A$ (as defined in Worthey \& Ottaviani 1997); and
the 4000\AA\ break (as defined in Balogh et al. 1999).

Quantitatively, we present least-squares fits to these data in
table~\ref{tab:fits}.  We show for each fit the slope and intercept as
well as Pearson's linear correlation coefficient $r$ (see e.g. Press
et al. 1992). Furthermore, we give an estimate labelled ``Pred.''  as
the predictability of each physical observable using the fit to
a given principal component. This number gives the RMS of the
distribution $(X_{\rm fit}-X_{\rm obs)/X_{\rm obs}}$, for every
physical observable $X$. Notice the strongest correlation appears
between PC3 and velocity dispersion, with a predictability of $\sim 4$\%
and between PC1,PC2 with colour, retrieving the observed colours
within a 10\% error. 

\begin{table*}
\begin{minipage}{15cm}
\caption{Linear fits of the Principal Component projections
to some physical observables}
\label{tab:fits}
\begin{center}
\begin{tabular}{cccccc}
\hline\hline
Y & X & Slope & Cut & r$^1$ & Pred.$^2$ \\
\hline
PC1$\times 10^3$ & $\log\sigma$ & $-5.071$ & $45.365$ & $-0.178$ & $0.206$ \\ 
PC2$\times 10^3$ & $\log\sigma$ & $-4.341$ & $9.827$ & $-0.426$ & $0.080$ \\ 
PC3$\times 10^3$ & $\log\sigma$ & $+2.033$ & $-4.624$ & $+0.685$ & $0.039$ \\ 
PC5$\times 10^3$ & $\log\sigma$ & $-0.339$ & $0.770$ & $-0.189$ & $0.197$ \\ 
\hline
PC1$\times 10^3$ & M$_r$ & $+0.214$ & $38.462$ & $+0.045$ & $0.522$ \\ 
PC2$\times 10^3$ & M$_r$ & $+0.303$ & $6.512$ & $+0.181$ & $0.131$ \\ 
PC3$\times 10^3$ & M$_r$ & $-0.201$ & $-4.347$ & $-0.411$ & $0.052$ \\ 
PC5$\times 10^3$ & M$_r$ & $+0.055$ & $1.199$ & $+0.188$ & $0.126$ \\ 
\hline
PC1$\times 10^3$ & $\log R_e$ & $-2.365$ & $35.004$ & $-0.030$ & $0.692$ \\ 
PC2$\times 10^3$ & $\log R_e$ & $-0.673$ & $0.286$ & $-0.024$ & $0.898$ \\ 
PC3$\times 10^3$ & $\log R_e$ & $+0.298$ & $-0.147$ & $+0.036$ & $0.578$ \\ 
PC5$\times 10^3$ & $\log R_e$ & $+0.027$ & $-0.014$ & $+0.005$ & $3.913$ \\ 
\hline
PC1$\times 10^3$ & $g-r$ & $-27.049$ & $54.287$ & $-0.570$ & $0.095$ \\ 
PC2$\times 10^3$ & $g-r$ & $-9.763$ & $7.335$ & $-0.578$ & $0.099$ \\ 
PC3$\times 10^3$ & $g-r$ & $+0.880$ & $-0.665$ & $+0.179$ & $0.384$ \\ 
PC5$\times 10^3$ & $g-r$ & $+0.367$ & $-0.278$ & $+0.123$ & $0.567$ \\ 
\hline
\hline
\end{tabular}
\end{center}
$^1$ Pearson's linear correlation coefficient (see e.g. Press et al. 1992, 
pg. 636).\\
$^2$ This is the RMS of the {\sl predictability} of the fit to
reproduce the physical observable $X$ from the measurement of the
component $Y$ (see text for details).
\end{minipage}
\end{table*}

\section{Comparison with a maximum likelihood method}

The advantage of PCA lies in its ability to extract the components
that hold most of the information (in the sense of variance) in a
model independent way. The principal components presented above do
not relate to any prescription regarding population synthesis or
galaxy formation. Unfortunately this also implies that in order
to put the ``physics'' back in, one has to compare the components
or the projections with models. Our main goal is
to relate the projections for our sample of galaxies with estimates of
age or metallicity of the underlying stellar populations.

We  compare our galaxies with a set of synthetic spectra using
the stellar population models of Bruzual \& Charlot (2003).  We
assume an exponentially decaying star formation rate (i.e. a $\tau$
model). Each star formation history (SFH) is described by three parameters,
namely the metallicity $Z$ -- kept fixed at all times; the time at which
star formation starts (which we can relate to a formation redshift $z_F$); 
and the timescale $\tau$ that controls the star formation rate. For each SFH 
we convolve the simple stellar populations -- using a Chabrier (2003) IMF --
into a composite spectrum
which is smoothed to the same velocity dispersion of the galaxy and 
sampled in the same way as the SDSS spectra. A $\chi^2$ is computed and
the parameters corresponding to a maximum likelihood is searched in 
a $24\times 24\times 24$ grid over a wide range of the parameters:
\begin{center}
$0.1\leq Z/Z_\odot\leq 2$\\
$0\leq\log(z_F)\leq 1$\\
$0.1\leq \tau /{\rm Gyr}\leq 6$.\\
\end{center}
\noindent
The best fit is subsequently improved by a Metropolis method where
extra models are randomly sampled and an accept/reject criterion based
on the likelihood is used to properly sample the probabilily
distribution of the parameters (see e.g. Saha 2003).  Here we use the
best fit values for the metallicity and the average age (which is a
function of $z_F$ and $\tau$). The best fits for these
galaxies have a mean $\langle\chi_r^2\rangle =0.9$ with RMS$=0.3$,
a typical sign of the degeneracies present in analyses of 
unresolved stellar populations. Synthetic models fit ``too well''
and it is very hard to disentangle these degeneracies.

\begin{figure}
\begin{center}
\includegraphics[width=5in]{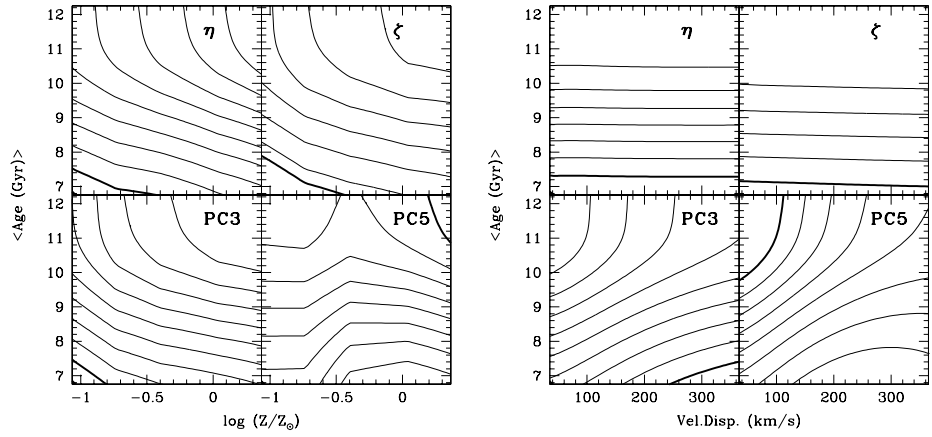}
\caption{
Synthetic spectra corresponding to an exponentially decaying star
formation history at fixed metallicity are built using the Bruzual \&
Charlot (2003) models and projected on the principal components
obtained for our sample. The figure shows contours of these
projections with respect to average age and either metallicity ({\sl
left}) or velocity dispersion ({\sl right}). The contour corresponding
to the highest value is shown as a thick line. Notice $\eta$ and
$\zeta$ (or similarly PC1 and PC2) are insensitive to velocity
dispersion, whereas PC3 is found to be strongly correlated with
velocity dispersion (see table~\ref{tab:fits}).}
\label{fig:contours}
\end{center}
\end{figure}

Figure~\ref{fig:contours} shows contours of the projections of the
model spectra on to the principal components. They are shown in the 2D
parameter space spanned by either average age and metallicity 
({\sl left}, assuming a fixed velocity dispersion of $\sigma=200$~km
s$^{-1}$), or average age and velocity dispersion ({\sl right},
assuming solar metallicity). The dashed line shows the contours with
the highest value of the projection in each case. The age-metallicity
degeneracy (see e.g. Worthey 1994) is apparent in the figure: one
cannot choose a set of parameters whose contours cross in a way to
univocally determine the value of age and metallicity. For instance,
$\eta$ and $\zeta$ have very similar dependence with age and
metallicity.  PC5 -- which features the characteristic Balmer
absorption lines -- appears to be more sensitive to age than
metallicity, a well-known fact exploited as the method to estimate
ages from Balmer indices (see e.g. Worthey \& Ottaviani 1997; Vazdekis
\& Arimoto 1999). Notice $\eta$ and $\zeta$ are insensitive to
velocity dispersion. The strong dependence of PC3 on the velocity
dispersion of the models shown in the right-hand panels of
figure~\ref{fig:contours} is consistent with the observed correlation
between PC3 and $\sigma$ shown in figure~\ref{fig:physobs}. The
observed correlation between PC1,PC2 and velocity dispersion
(table~\ref{tab:fits}) is caused by the intrinsic correlation between
$\sigma$ and colour (see e.g. Bernardi et al. 2003).

\begin{figure*}
\includegraphics[width=16cm]{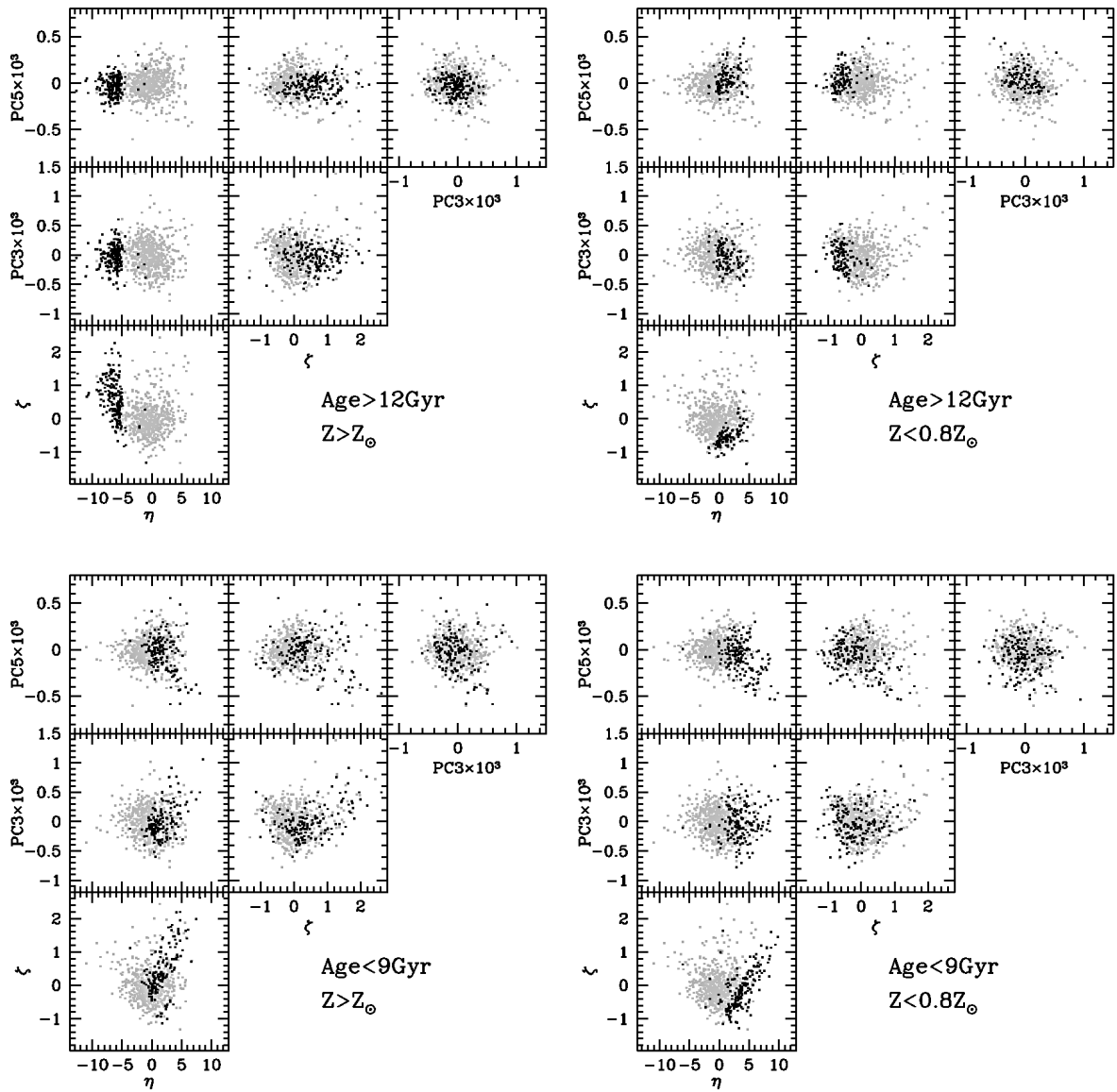}
\caption{Comparison of the principal component projections with
age/metallicity estimates using a direct fit to the full spectra via a
$\chi^2$ comparison with composite stellar populations from Bruzual \&
Charlot (2003) using an exponentially decaying star formation rate
(see text for details). The grey dots show the total sample. The
black dots correspond to subsamples which are best fit by models
whose range of ages and metallicites are given in each panel.}
\label{fig:chi2}
\end{figure*}

Figure~\ref{fig:chi2} compares subsets of galaxies with a given age
and metallicity -- estimated by the maximum likelihood method
described above -- in the parameter space spanned by our principal
components. The figure show as grey dots the projections of the total
sample and as black dots those galaxies corresponding to different age
and metallicity distributions. The top panels correspond to old ages
($\langle t\rangle\geq 12$Gyr) and the bottom panels are best fit by
younger populations ($\langle t\rangle\leq 9$Gyr). Left and right
panels are separated with respect to metallicity, as labelled.

Notice that the dependence of $\eta$ and $\zeta$ with respect
to metallicity is very different for old and young populations. As we
change the metallicity of young stars, the $\zeta$ parameter increases
without a strong change in $\eta$, whereas a range in metallicity in
old stellar populations has an effect both in $\eta$ and $\zeta$. This
will be an important trend to bear in mind when UV fluxes are taken
into account (in the next section).

\begin{figure*}
\includegraphics[width=16cm]{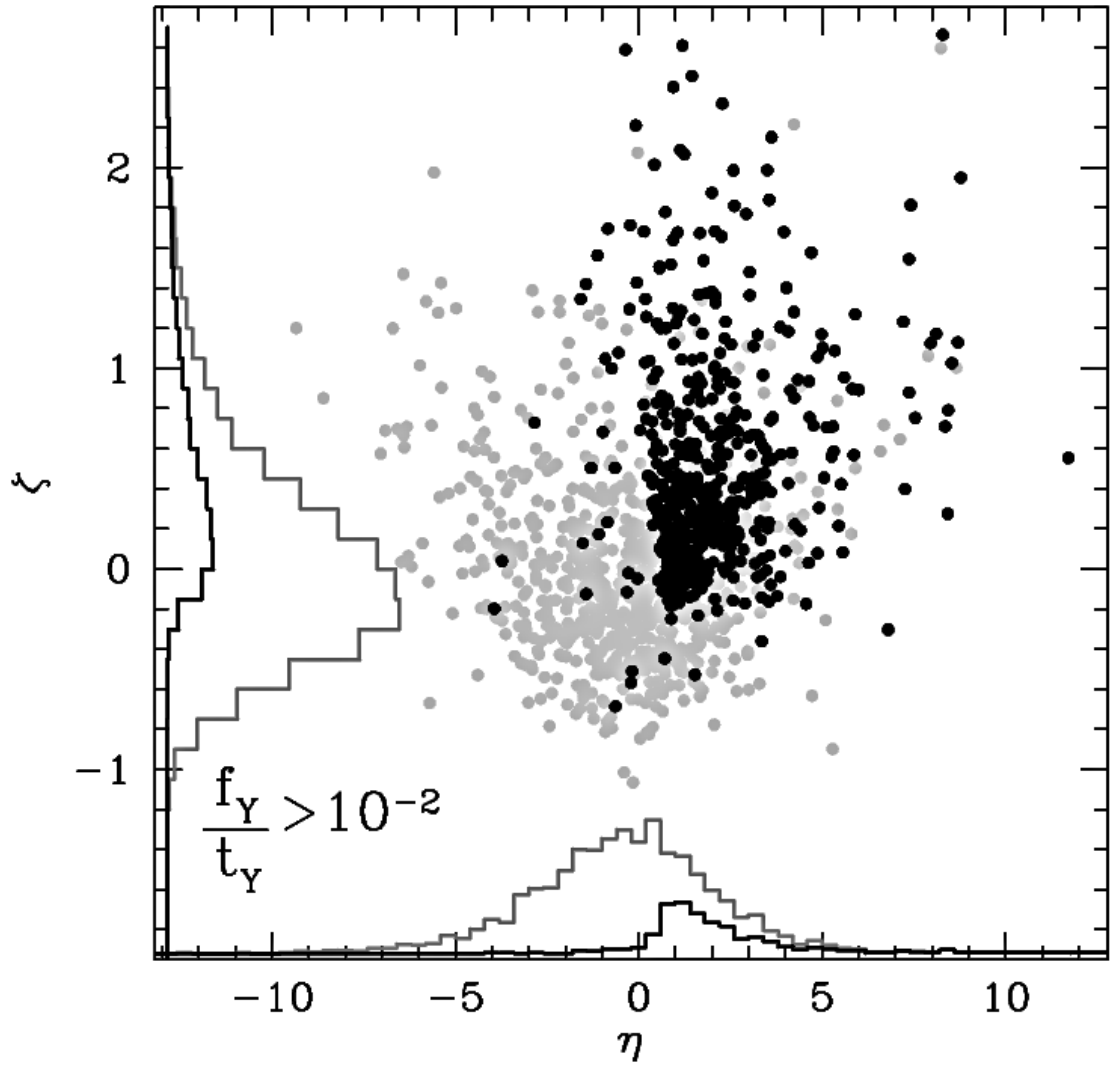}
\caption{Comparison of the principal component projections with a 
model that overlays a young stellar population (age $t_Y$; mass
fraction $f_Y$) on top of an old component (see text for details).
The whole sample is shown as grey dots. The black dots correspond to
galaxies whose best fit requires $f_Y/t_Y >10^{-2}$ (e.g.  1\% in
1~Gyr old stars). The histogram in the $\eta$ and $\zeta$ parameters
are shown for both sets of galaxies with the same colour coding.}
\label{fig:2bc2}
\end{figure*}

An alternative set of models to compare with the spectra consist
of a young (simple) stellar population added to an overall old
component. This two-burst method is motivated by the idea of
``frostings'' of young stars over an otherwise old population (Trager
et al. 2000) and has been succesfully used to explore recent star
formation in elliptical galaxies (Ferreras \& Silk 2000; Kaviraj et
al. 2007; Schawinski et al. 2006).  Figure~\ref{fig:2bc2} shows a
comparison with these models.  The old component is a $\tau$-model with a
formation redshift $z_F=5$ and star formation timescale
$\tau=1$~Gyr. Three parameters are explored to find the best fit: the
age (t$_Y$) and mass fraction (f$_Y$) of the young component; and the
metallicity of the galaxy  --which is assumed to be the same for both
old and young components to reduce the number of free
parameters. Using the same technique as described above, we show in
the figure the results for the $\eta$ and $\zeta$ projections. The
grey dots correspond to the whole sample, whereas we show as black
dots those galaxies whose spectral fitting requires a significant
``frosting''. We chose those with $f_Y/t_Y>0.01$, which implies e.g. 
a $1$\% contribution in 1~Gyr stars or a $0.1$\% contribution in 100~Myr
stars. These galaxies populate the region corresponding to a higher
value of $\eta$ and $\zeta$. We emphasize here that these are just
simple models to guide our analysis based on the principal
components. The real galaxies -- which we will explore in more detail
in the next section -- have a distribution of stellar populations much
more complicated than any of the simple models described here (or
elsewhere).  Hence the advantage of multivariate, model independent,
analysis.

\section{Comparison with Near Ultraviolet Photometry}

Photons detected from early-type galaxies in the ultraviolet spectral
window come from two main sources: evolved Horizontal Branch stars and
their progeny or recent episodes of star formation (O'Connell 1999).
The former are responsible for the UV upturn around 1200\AA\ and it
was not clear until recently whether near ultraviolet (NUV) light 
could be only attributed to evolved stars in early-type galaxies. The
recent survey of SDSS early-type galaxies targeted by GALEX (Yi et
al. 2005) confirmed that a small albeit measurable
recent star formation (RSF) is present in a significant number of early-type
galaxies. The observed flux in the near ultraviolet passband of GALEX
centered at 2350\AA\ is most sensitive to hot dwarfs and even though
there can be some contribution from evolved HB stars, the NUV$-r$
colours observed in many of the targeted ellipticals cannot be
explained purely by an old population. Schawinski et al. (2006)
estimate that galaxies with colours NUV$-r<5.4$ must have a fraction
of young stars. This fraction can be as small as a few percent
(Kaviraj et al. 2007) and it is not clear what causes this episode of
star formation.

Our sample of $7148$ galaxies includes $921$ objects that overlap with
the GALEX Medium Imaging Survey (Morrissey et al. 2005). For details
on the photometry we refer to Schawinski et al. (2006) and Kaviraj et
al. (2007).  In order to enhance the contrast between NUV bright and
NUV faint galaxies, we select those with NUV$-r\leq 4.9$ as blue
galaxies and NUV$-r\geq 5.9$ as red galaxies. The subsamples comprise
$151$ blue and $241$ red galaxies. The mean redshift of both
subsamples is $\langle z\rangle = 0.076$, compatible with the average
redshift of our sample ($0.075$). Figure~\ref{fig:Galex} shows the
histogram of the projections of these two subsamples on our principal
components. The distribution of $\eta$, PC3 and PC5 are
indistinguishable, whereas the histogram corresponding to $\zeta$
features a significant skew towards higher values for the blue
subsample. In the light of the comparison with synthetic models it is
remarkable to find such a result given that the estimated amount of
recent star formation stays below a few percent (Kaviraj et al. 2007).
Furthermore, instead of NUV, we are using the less sensitive optical
spectral window to detect such effect.  This illustrates the power of
PCA to extract small differences from the spectral energy
distribution. This result allows us to further relate the $\zeta$
parameter to recent star formation.  Notice that
figure~\ref{fig:contours} show that {\sl both} $\eta$ and $\zeta$
depend on age (and metallicity), whereas the comparison with GALEX
data suggest $\zeta$ to be the only parameter sensitive to recent star
formation.

The bottom panels of figure~\ref{fig:Galex} show the histograms of PC1 and
PC2 for the continuum-subtracted spectra. The separation is much harder
to determine, however a slight difference can be seen in the histograms
of PC2 (in analogy with $\zeta$ for the full SED case). Even though this
result is very weak, it is quite remarkable given that the information
regarding colours is completely eliminated when subtracting the continuum.

The lines in the top panels correspond to simple models overlaying a
young (0.5 Gyr) population over an old component (12 Gyr), for three
different metallicities as labelled. The lines correspond to a change
in $f_Y$, the stellar mass fraction in young stars. Notice the difference
in the predicted evolution of $\eta$ and $\zeta$ as $f_Y$ is increased.
The lines illustrate that a small ``frosting'' (as in Trager et al. 2000)
in young, metal-rich stars can increase the value of $\zeta$ without a large
change in $\eta$. Hence $\zeta$ is the parameter most sensitive to
recent (and small) episodes of star formation as long as the metallicity
is high. More intense star formation stages (or at lower metallicity)
would reflect in changes both for $\eta$ and $\zeta$.

\section{Comparison with the Composite Spectra of Bernardi et al.}

In order to compare our results with the more established method
focussing on the equivalent widths of some absorption lines
with strong age or metallicity trends (e.g. Worthey \& Ottaviani 1997;
Kuntschner \& Davies 1998; Vazdekis \& Arimoto 1999; Trager et al. 2000;  
Thomas \& Maraston 2003; Bernardi et al. 2003) we apply the composite spectra
of Bernardi et al. (2006).
These models were generated from the original sample from which
our catalogue is extracted. Galaxies were arranged in bins
of redshift, absolute magnitude, velocity dispersion, etc... and
combined to generate composite SEDs with a S/N high enough for
an analysis of equivalent widths. We refer the reader to Bernardi 
et al. (2006) for a detailed study of these composite spectra.

\begin{figure}
\begin{center}
\includegraphics[width=15cm]{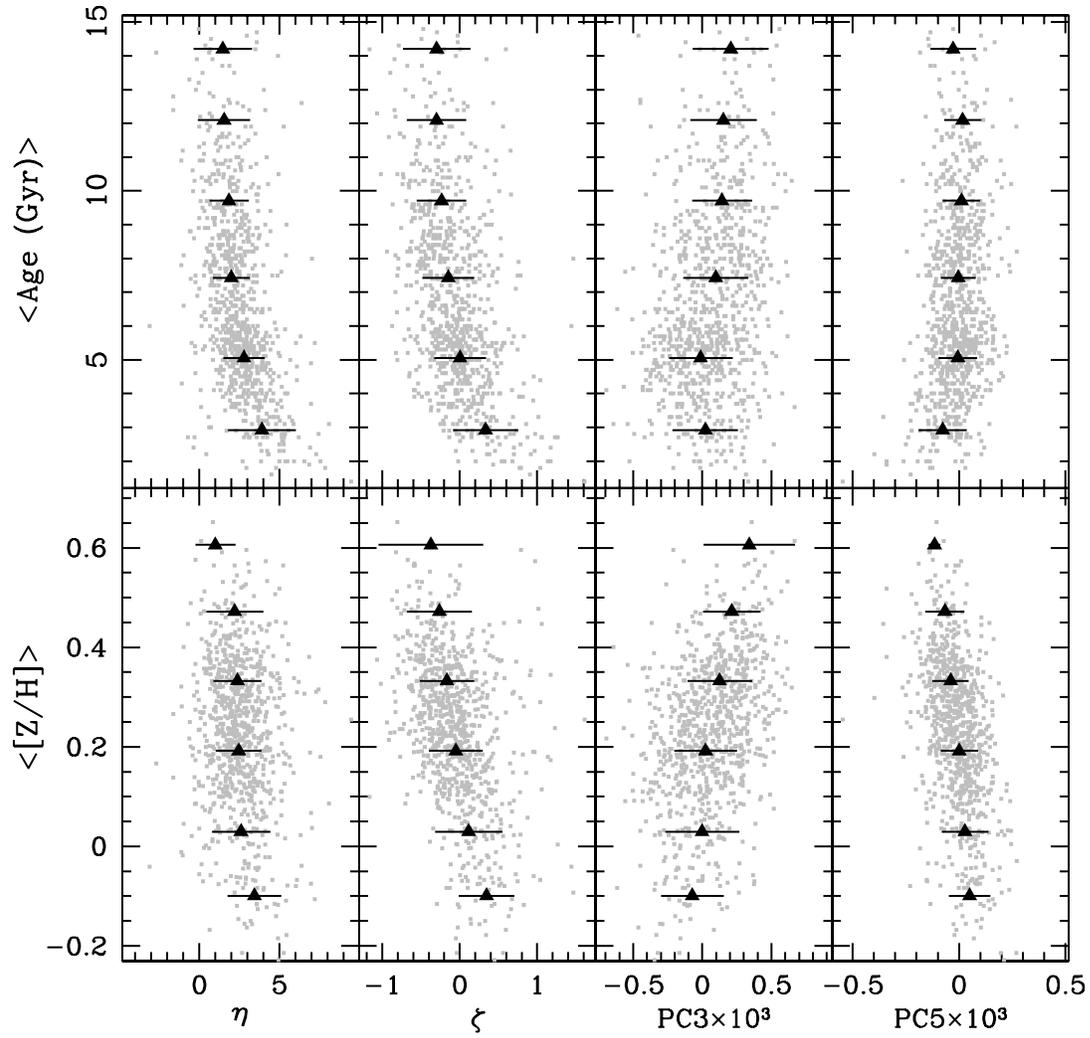}
\end{center}
\caption{The composite spectra of Bernardi et al. (2006) are
projected on our principal components. The age and metallicity
estimates are presented in Bernardi et al. (2006) and are
computed using a complete analysis of Lick indices and the
H$\gamma$ absorption line. The triangles and error bars
are average and standard deviation of data binned either
in age or metallicity.}
\label{fig:BerMods}
\end{figure}

Figure~\ref{fig:BerMods} shows a comparison between the projections
of the composite SEDs on our principal components and the age and
metallicity estimates obtained from an analysis of spectral lines
including H$\gamma_A$. Only those models with our redshift ($z\leq 0.1$) 
and absolute magnitude ($M_r<-21$) limits were considered. The triangles
and error bars give the mean and standard deviation of data binned
in age or metallicity. The figure shows the dependence of 
the projections on both age and metallicity. High values of 
$\eta$ and $\zeta$ correspond to younger stellar ages. However,
the interplay between age and metallicity is complicated and one
can find -- consistently with figure~\ref{fig:chi2} -- that a 
high value of $\eta$ can be a signature of younger ages or lower metallicities.
Only when $\eta$ changes without a strong change in $\zeta$ can
we disentangle the effects and associate the SED with recent star formation
(as in the NUV bright/faint separation shown above).

\section{Discussion and conclusions}

\begin{figure*}
\includegraphics[width=15cm]{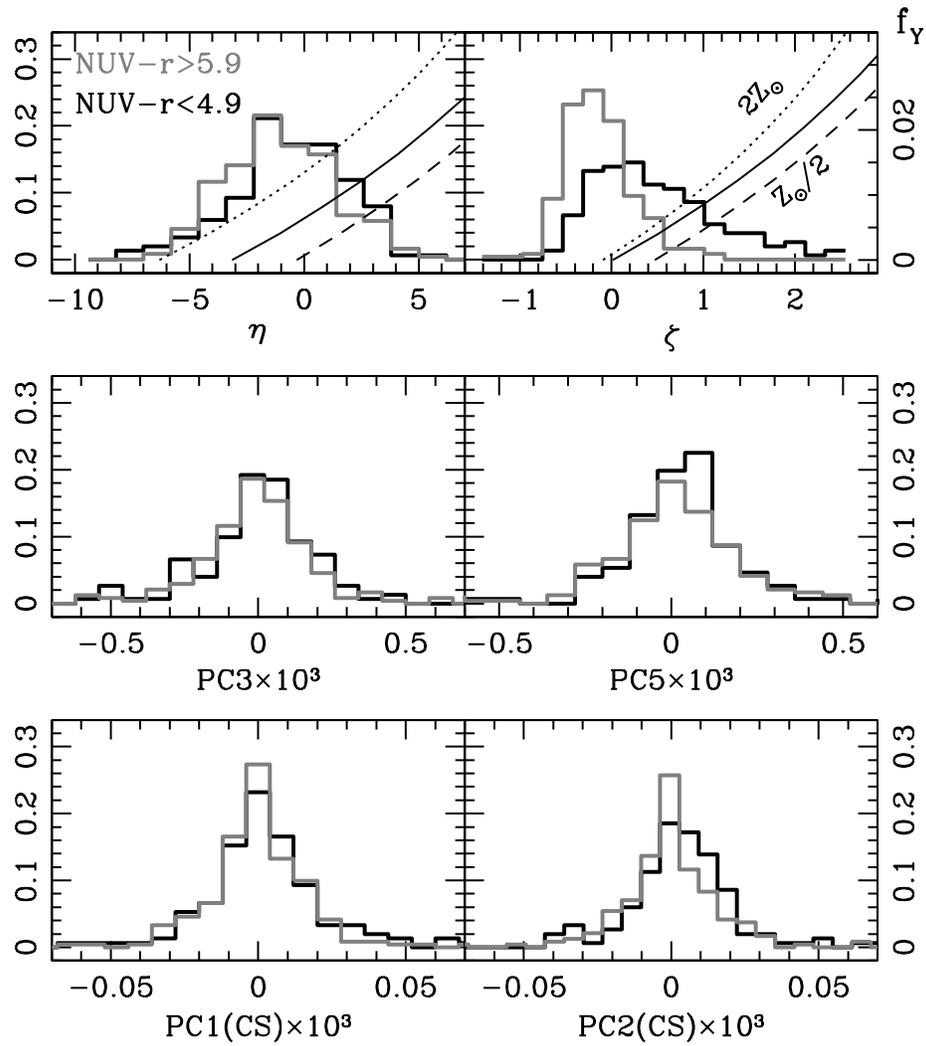}
\caption{Distribution of the projections of the galaxy
spectra on to our principal components. A subsample with
available GALEX NUV data is chosen, and split according
to NUV$-r$ colour. The NUV bright galaxies (black histograms)
are expected to have undergone an episode of recent star formation
(Schawinski et al. 2006). These galaxies have a different
distribution of $\zeta$ but the distribution of the other
components is indistinguishable from the red subsample. The
bottom panels show the distribution of the projections of the
first and second principal components for the continuum-subtracted
case. Although small, there is a difference in the histogram
of PC2 (which mainly relates to $\zeta$ for PCA with the full SEDs).}
\label{fig:Galex}
\end{figure*}

With the comparison between the projections of the galaxy spectra on
to the principal components and their relationship with physical
observables such as those obtained from synthetic spectra (\S4) or NUV
photometry (\S5) we can put some physics back into the analysis.  PCA
reduces most of the information -- in the sense of variance -- to a
biparametric set defined by $\eta$ and $\zeta$. This result is
reminiscent of the work of Faber (1973). We have found that a
variation of the $\zeta$ component can disentangle the effect of small
amounts of young star formation as long as the young burst corresponds
to high metallicities. This effect is apparent in a comparison of the
distributions of the $\eta$ and $\zeta$ projections for NUV bright and
faint galaxies as in figure~\ref{fig:Galex}.

No significant effect is found regarding the fact that our
spectra are extracted from a fibre, which will map different physical
sizes at different redshifts. From the lowest redshift found in our
sample ($z=0.03$) to the $z=0.1$ limit imposed, the $3^{\prime\prime}$
diameter of the fibre covers a physical size from $1.8$ to
$5.5$~kpc. However, the steep surface brightness profile of early-type
systems and the not-too-deep exposure of the fibres implies that most
of the light from the observed spectra comes from the inner regions of
these galaxies regardless of redshift.  We compared the projections of
the principal components $\eta$ and $\zeta$ as well as the higher
order ones, to find no significant trend with redshift. 
For instance, the difference in the distribution of $\eta$ and $\zeta$
found between NUV-bright and faint galaxies -- shown in
figure~\ref{fig:Galex} -- is preserved when subsamples with redshift only
below (or above) the median redshift are chosen. Hence we conclude our
analysis is insensitive to aperture effects.

\begin{figure}
\begin{center}
\includegraphics[width=15cm]{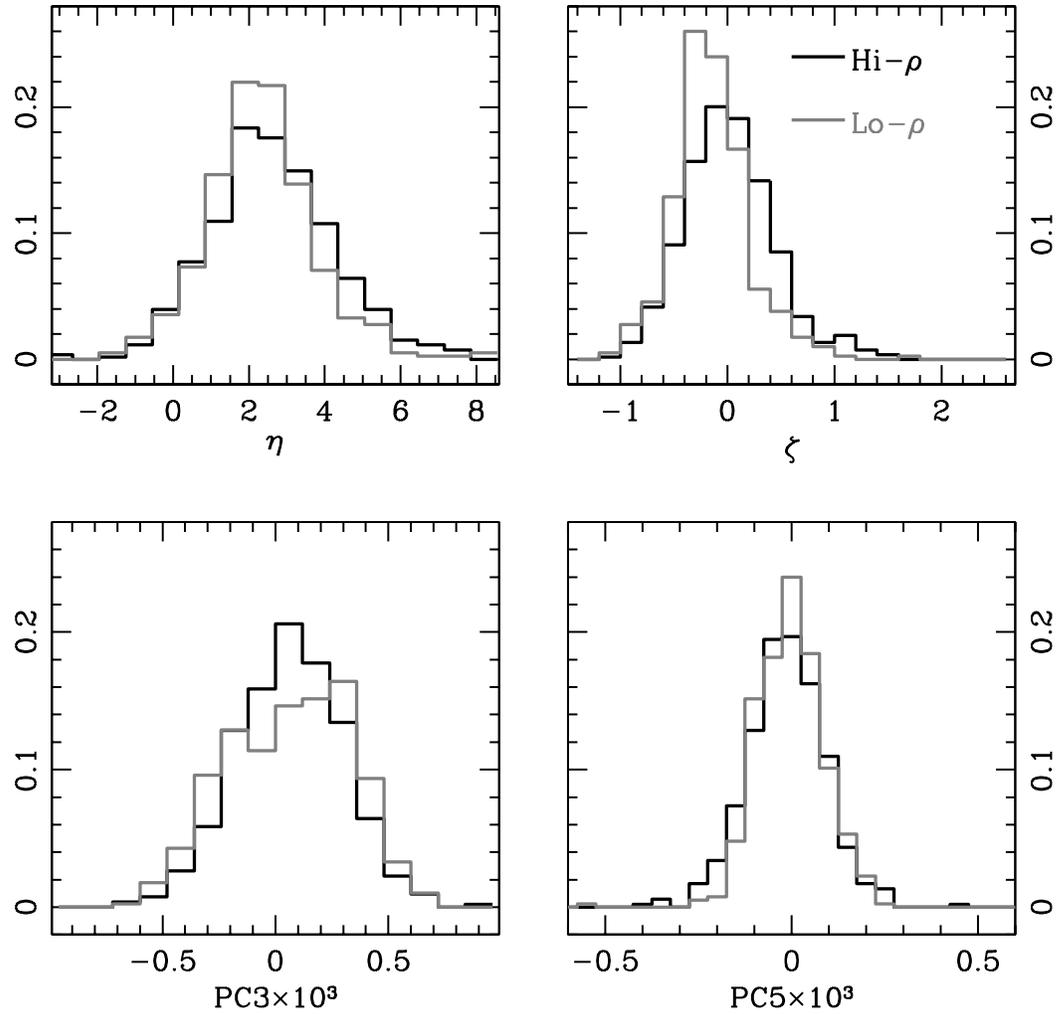}
\end{center}
\caption{Comparison of the projections of the composite spectra on the
principal components with respect to environment -- as defined in 
Bernardi el al. (2006). Analogous to the results comparing distirbutions
binned with respect to NUV$-r$
colours, the difference between galaxies in low and high density regions
is found in the $\zeta$ parameter, with a skew towards high values
for galaxies living in high-density regions.}
\label{fig:dens}
\end{figure}

We can extend this analysis to the composite spectra of Bernardi et
al. (2006). Among various properties, these models are binned
according to average density. Hence, we can separate these composite
models into high- and low-density galaxies depending on whether the
distance to the nearest cluster {\sl and} the distance to the 10th
nearest neighbour are both smaller or greater than 10~Mpc,
respectively. The distributions of the projections of the high/low
density composite spectra on the principal components are shown in
figure~\ref{fig:dens}.  In parallel to the comparison with NUV$-r$
colours (figure~\ref{fig:Galex}) we find no significant difference in
the distributions of $\eta$, PC3 and PC5, 
whereas $\zeta$ presents a skew towards higher values for
galaxies in our higher density regions {\sl analogous to NUV$-r$ blue
galaxies}. Our ``high-density bin''
includes galaxies across a wide range of environments: from 
groups to clusters: both central regions and outskirts.  In this paper
we show that early-type galaxies in {\sl low density regions} do NOT
present those signs of recent star formation, compared to their
counterparts in higher density regions (cf. Denicol\'o et al. 2006;
Clemens et al. 2006; Schawinski et al. 2006; de la Rosa et
al. 2007). An ongoing project will explore the environmental issue in more
detail (Ferreras et al., in preparation). Nevertheless, from the data
presented in this paper we can safely conclude that environmental
effects are very small in early-type galaxies: PCA finds a departure
from a homogeneous stellar population across different environments of
less than 1\%.

\begin{figure}
\begin{center}
\includegraphics[width=15cm]{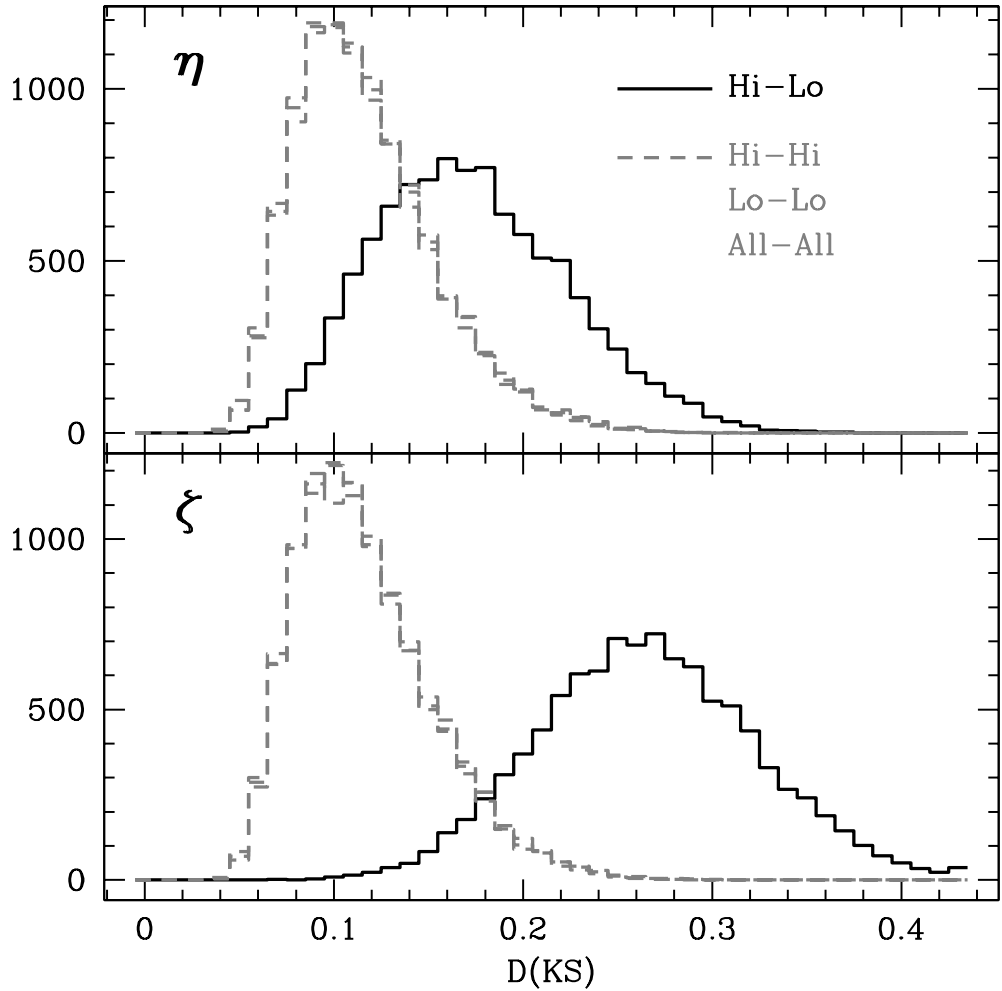}
\end{center}
\caption{Checking the difference found between elliptical galaxies in high and
low-density regions. The solid histograms show the distribution of the
D-statistic from a Kolmogorov-Smirnov test applied to 10000
realisations on subsamples of 100 spectra each, extracted from the
high and low-density samples. The dashed line histograms are similar
distributions when extracted from the same density samples, either
high; low-density or from the complete sample of composites.  }
\label{fig:dens2}
\end{figure}

These results confirm the minor effect that the environment seems to
play on the star formation histories of early-type galaxies, in
agreement with previous work (see e.g. Bernardi et al. 1998; Smith et
al. 2006). Nevertheless, the difference between the high and the
low-density distributions is significant: A Kolmogorov-Smirnov test
comparing the 268 high-density composites and the 314 low-density
composites give a KS statistic of D=0.145 and 0.237 for $\eta$ and
$\zeta$, respectively, implying a statistical significance of over
99\% that both samples do NOT come from the same distribution.  In
order to strengthen this point, we show in figure~\ref{fig:dens2} a
Monte Carlo test.  We perform 10000 comparisons between subsamples of
100 spectra each.  The solid line shows the distribution of the
D-statistic from the Kolmogorov-Smirnov test when applied between
subsamples in high and low-density regions, whereas the grey, dashed
line histograms correspond to comparisons between subsamples obtained
from the same density bin, or from the whole set of composites.  Both
$\eta$ and $\zeta$ are clearly split when comparing subsamples in
different density regions, but the latter shows a very strong
separation.

The interpretation of the mild correlation would suggest galaxies
in higher density regions (most likely groups and the outskirts of
clusters) to have weak episodes of recent star formation at high
metallicity. These high metallicities are to be expected in the
heavily polluted interstellar medium of an early-type galaxy.  This
result is consistent with a recent analysis of a smaller sample
comprising early-types both in Hickson Compact Groups and in the field
(Ferreras et al. 2006). In this sample, galaxies in groups present a
stronger second principal component compared to elliptical galaxies in
the field, i.e. frostings of recent star formation. It is
interesting to contrast this result with Nolan et al. (2007) who find
E+A 'post-starburst' galaxies (not-necessarily with an early-type
morphology) mainly in group/field environments, with a significant
decrease at higher densities. In combination with the analysis
presented in this paper, the data hint at a peak in the (frostings)
activity of early-type galaxies at group densities.  The weakness of
these star formation episodes is also consistent with the small amount
of cold gas that could survive in this type of galaxies (Faber \&
Gallagher 1976).  There is recent evidence from a CO emission survey
targeting SAURON early-type galaxies which suggests that these
episodes of star formation should be fuelled by the infall of small
amounts of gas from the outside (Combes, Young \& Bureau 2007). The
fraction in young stars is estimated to be a few percent, which is so
small as to challenge model-dependent methods.

\section*{Acknowledgements}

BR gratefully acknowledges support from the Royal Astronomical
Society. IF acknowledges a grant from the Nuffield Foundation. 
This work was supported by grant No. R01-2006-000-10716-0 from 
the Basic Research Program of {\sl KOSEF} to SKY. We
would like to thank the anonymous referee for her/his very interesting
comments and suggestions.  Funding for the SDSS and SDSS-II has been
provided by the Alfred P. Sloan Foundation, the Participating
Institutions, the National Science Foundation, the U.S. Department of
Energy, the National Aeronautics and Space Administration, the
Japanese Monbukagakusho, the Max Planck Society, and the Higher
Education Funding Council for England. The SDSS Web Site is
http://www.sdss.org/.  The SDSS is managed by the Astrophysical
Research Consortium for the Participating Institutions.


\onecolumn

\end{document}